\documentclass{article}
\pdfoutput=1

\usepackage{listings}

\usepackage{xcolor}

\usepackage{graphicx}

\usepackage{subfig}

\usepackage{xspace}

\usepackage{todonotes}

\usepackage{tikz}

\usepackage{amsmath,amssymb}

\newcommand*\circled[2][black]{\tikz[baseline=(char.base)]{
            \node[shape=circle,fill,inner sep=2pt,fill=#1] (char) {\textcolor{white}{#2}};}}

\makeatletter
\DeclareRobustCommand\onedot{\futurelet\@let@token\@onedot}
\def\@onedot{\ifx\@let@token.\else.\null\fi\xspace}

\def\eg{\emph{e.g}\onedot} 
\def\ie{\emph{i.e}\onedot} 
 
\def\etc{\emph{etc}\onedot}

\makeatother

\graphicspath{ {./figures/} }

\definecolor{codegreen}{rgb}{0,0.6,0}
\definecolor{codegray}{rgb}{0.5,0.5,0.5}
\definecolor{codepurple}{rgb}{0.58,0,0.82}
\definecolor{backcolour}{rgb}{0.95,0.95,0.92}
\definecolor{deepblue}{rgb}{0,0,0.5}
\definecolor{deepred}{rgb}{0.6,0,0}
\definecolor{deepgreen}{rgb}{0,0.5,0}

\lstdefinestyle{mystyle2}{
  language=Python, 
  morekeywords={search,rule,bm25_match,neural_match,neural_extract,op_filter,bm,_match,translate,answer,eval,fact},
  otherkeywords={0,1,2,3,4,5,6,7,8,9,=,<,>,ids(),vars()},
  backgroundcolor=\color{white},%
  commentstyle=\color{codegreen},
  keywordstyle=\color{codepurple},%
  numberstyle=\tiny\color{codegray},
  stringstyle=\color{deepred},%
  basicstyle=\ttfamily\normalsize,%
  emph={_asin , _total_reviews , _price , _title , _review , _review_text , _answers, NeuroQL},     
  emphstyle=\color{blue}, 
  breakatwhitespace=false,         
  captionpos=b,                    
  keepspaces=true,                 
  numbers=left,                    
  numbersep=5pt,                  
  showspaces=false,                
  showstringspaces=false,
  showtabs=false,                  
  tabsize=2,
}

\lstdefinestyle{datastyle}{
  language=Python, 
  backgroundcolor=\color{white},%
  commentstyle=\color{codegreen},
  keywordstyle=\color{codepurple},%
  otherkeywords={title,price,brand,stars,total_,total_reviews,review,item_weight,item_weight_units,item_model_number,is_discontinued_by_manufacturer,?asin,?title,?price,ids,search,==,translate,answer,answers,eval,fact},
  numberstyle=\tiny\color{codegray}, 
  stringstyle=\color{deepred},%
  basicstyle=\ttfamily\normalsize,%
  emph={B00001P4ZH, B000AJIF4E},     
  breakatwhitespace=false,         
  captionpos=b,                    
  keepspaces=true,                 
  numbers=left,                    
  numbersep=5pt,                  
  showspaces=false,                
  showstringspaces=false,
  showtabs=false,                  
  tabsize=2
}

\PassOptionsToPackage{numbers, compress}{natbib}

\usepackage[preprint]{neurips_2020}
\usepackage[utf8]{inputenc} %
\usepackage[T1]{fontenc}    %
\usepackage[colorlinks=true,allcolors=blue]{hyperref}       %
\usepackage{url}            %
\usepackage{booktabs}       %
\usepackage{amsfonts}       %
\usepackage{nicefrac}       %
\usepackage{microtype}      %

\title{NeuroQL: A Neuro-Symbolic Language and Dataset for Inter-Subjective Reasoning}

\author{%
  Nick Papoulias\\
  Director of Research\\
  OrgD. Labs\\
  \url{https://orgdlabs.com} \\
  \texttt{npapoylias@orgdlabs.com} \\
}

\begin{document}

\maketitle

\begin{abstract}
  We present a new AI task and baseline solution for Inter-Subjective Reasoning. We define inter-subjective information,  to be a mixture of objective and subjective information possibly shared by different parties. Examples may include commodities and their objective properties as reported by IR (Information Retrieval) systems, that need to be cross-referenced with subjective user reviews from an online forum. For an AI system to successfully reason about both, it needs to be able to combine symbolic reasoning of objective facts with the shared consensus found on subjective user reviews. To this end we introduce the NeuroQL dataset and DSL (Domain-specific Language) as a baseline solution for this problem. NeuroQL is a neuro-symbolic language that extends logical unification with neural primitives for extraction and retrieval. It can function as a target for automatic translation of inter-subjective questions (posed in natural language) into the neuro-symbolic code that can answer them.
\end{abstract}
 
\section{Introduction}\label{sec:intro}

Our digital world comprises of information sources with varying degrees of objectivity and subjectivity. Structured
information stored in databases or other IR systems (such as for \eg the physical properties and prices of products) are presented as objective facts to potential customers, while their accompanying free-form reviews  fall by definition in the subjective part of this spectrum. Yet, as users of these systems we are interested in questions involving both aspects. These are questions where both the authoritative structured data of a product and the unstructured opinions of the general public are pertinent. Consider for example the search query: 
\textit{``How is the bass for headphones at around 30 dollars having minimum 14K reviews that is not discontinued?''} which we depict in Fig. \ref{fig:question}: 

\begin{figure}[h!]
  \centering
  \advance\leftskip-1.4cm 
  \includegraphics[width=1.2\textwidth]{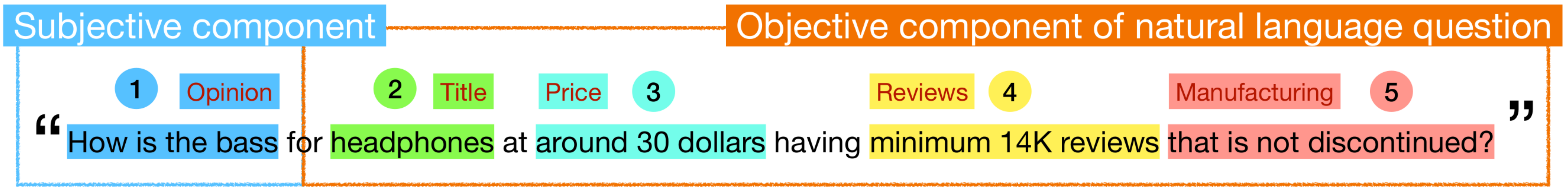}
  \caption{An inter-subjective question \textit{``How is the bass for headphones at around 30 dollars having minimum 14K reviews that is not discontinued?''}. Depicting both the subjective and the objective components of this question: (\ie \textit{opinion, title, price, reviews, manufacturing} \etc).}
  \label{fig:question}
\end{figure} 

We call this type of questions \textit{inter-subjective} given that they involve a mixture of subjective and objective information that may be shared by different parties. Currently, answers to such queries are treated with a semi-manual two stage process by information systems. In the first step the user needs to pinpoint a single product of interest using a query involving only its objective properties and description (e.g a query involving the title, price, number of reviews \etc as in the example above). Then in the second step, the user needs to scan the reviews of this product either manually or with a similarity search in order to find relevant subjective opinions regarding the quality of the product (e.g. the quality of the bass in the above example). With the advent of deep-learning \cite{lecun2015deep,goodfellow2016deep} for natural-language processing \cite{vaswani2017attention} and more specifically deep-learning Q\&A (Question \& Answering) models \cite{tunstall2022natural}, this second step can now be supplemented with neural retrieval and neural comprehension of reviews. 

In this work we investigate the possibility of automating and merging these two stages that involve both symbolic reasoning (over structured factual information) and neuronal reasoning (over unstructured subjective sources). The problem of interfacing symbolic and neuronal reasoning, is a known open problem in AI literature involving neuro-symbolic systems \cite{chaudhuri2021neurosymbolic,nye2021improving}. Our contribution is to propose a new AI task and baseline solution for \textit{inter-subjective} queries and reasoning, as the one we saw above. 

To this end we introduce the NeuroQL dataset and DSL (Domain-specific Language \cite{fowler2010domain,papoulias2019lands}) as a baseline solution for this problem. Our dataset extends previous work on Q\&A systems focused on metadata \cite{ni2019justifying} and subjective \cite{bjerva2020subjqa} information to include inter-subjective questions and their translation into neuro-symbolic queries.  Our queries are expressed in NeuroQL which is a neuro-symbolic language that we embed inside the Python \cite{van2023python} runtime environment. This embedded DSL implements and extends logical unification \cite{robinson1965machine,robinson2001handbook} with neural primitives for extraction and retrieval. NeuroQL can function as a target for automatic translation of inter-subjective questions (posed in natural language) into the neuro-symbolic code that can answer them, as we show in Figure \ref{fig:translation}: 
\vspace{-1.1em}
\begin{figure}[h!] 
  \centering
  \advance\leftskip-2.4cm 
  \includegraphics[width=1.35\textwidth]{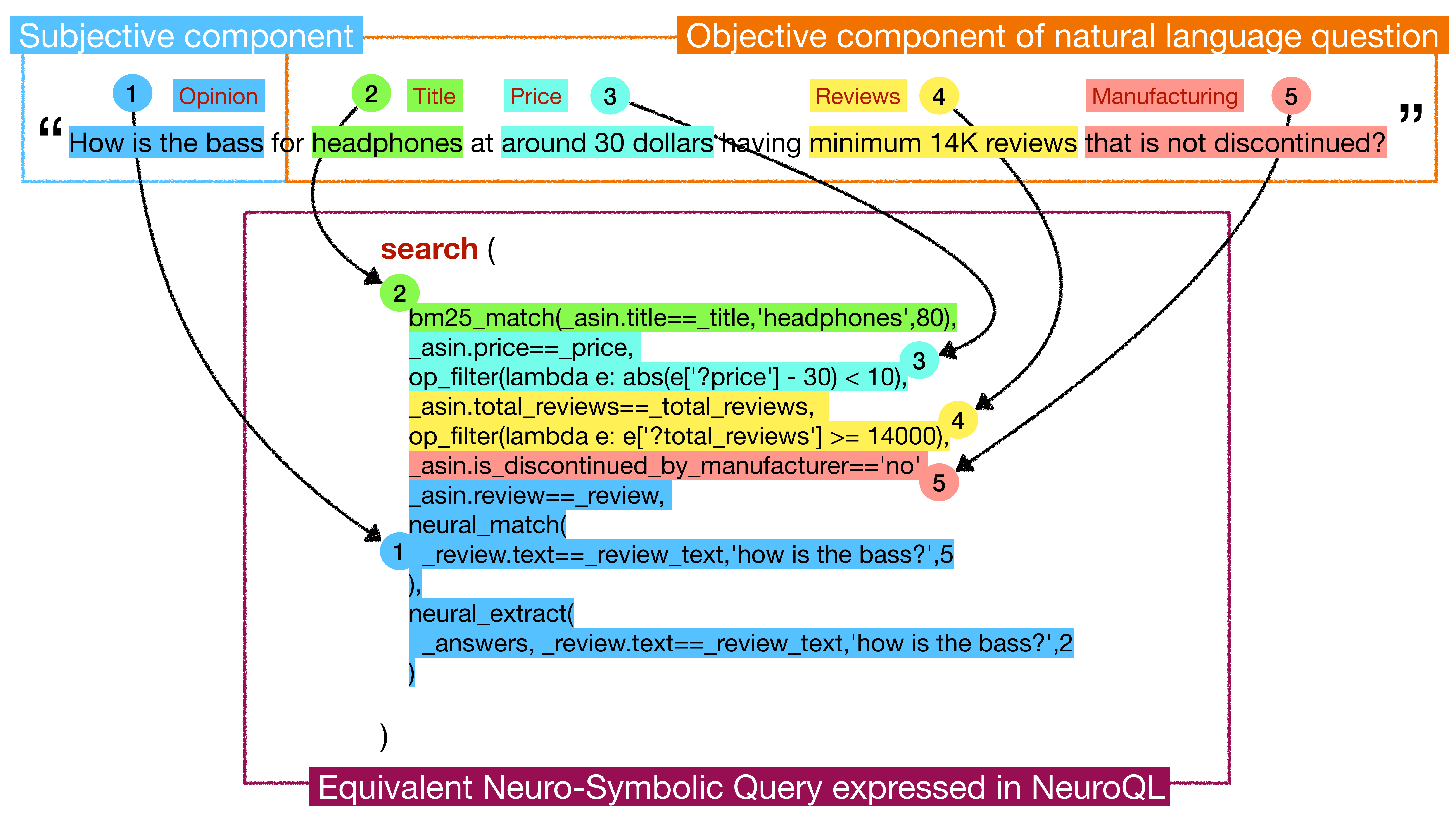}
  \caption{Equivalence between the inter-subjective question \textit{``How is the bass for headphones at around 30 dollars having minimum 14K reviews that is not discontinued?''} and its translation into a neuro-symbolic query in NeuroQL. Starting from the top we depict both the subjective and the objective components of the question and their corresponding categories (\ie \textit{opinion, title, price, reviews, manufacturing} \etc). On the bottom we see the corresponding \textit{search} expression in NeuroQL with all sub-queries highlighted to match their equivalent category in natural language.}
  \label{fig:translation}
\end{figure}

In this figure we show our goals and hypotheses regarding NeuroQL. Firstly, we hypothesize  that: 
\begin{quote}
  (\textbf{\textit{H1}}): \textit{It is possible to translate inter-subjective questions (as the one shown on top of Figure \ref{fig:translation}), comprising of different subjective and objective components, into neuro-symbolic queries expressed in NeuroQL.} 
\end{quote}

To test (\textbf{\textit{H1}}) we experiment with a neural translation solution fine-tuned to this domain, aiming to distinguish not only between objective and subjective components, but also between different kinds of sub-query categories (\ie \textit{opinion, title, price, reviews, manufacturing} \etc). These sub-query categories have different translations in NeuroQL (as shown in the bottom part of Figure \ref{fig:translation}) which are highlighted to match their equivalent categories in natural language. 

Furthermore, we hypothesize that: 

\begin{quote} 
  (\textbf{\textit{H2}}): \textit{Symbolic reasoning through unification (covering structured objective facts), similar to the one found in prolog \cite{clocksin2003programming,colmerauer1990introduction} and datalog \cite{ceri1989you} systems, can be extended with neuronal reasoning (for unstructured subjective data) in a disciplined manner, producing satisfactory answers for inter-subjective queries.}
\end{quote}  

Meaning, that the neuro-symbolic synthesis can be expressed in a concise syntactic and semantic form for NeuroQL users, while still being able to answer inter-subjective queries in a satisfactory way. To test (\textbf{\textit{H2}}) we measure \textit{recall, em and f1 scores} \cite{tunstall2022natural} of translated NeuroQL queries against previously unseen inter-subjective questions.

The rest of this paper is organized as follows: Section \ref{sec:results} presents the results of our effort including: a hands-on walk-through of NeuroQL (sub-section \ref{sec:neuroql_by_example}), a description of the NeuroQL inference pipeline (sub-section \ref{sec:neuroql_architecture}), a description of the NeuroQL dataset (sub-section \ref{sec:neuroql_dataset}), our translation, recall, em and f1 experiments (sub-section \ref{sec:experiments_and_validation}) as well as a further review of related work (sub-section \ref{sec:related_work}). Section \ref{sec:discussion} discusses the implications of our contribution, possible threats to validity and future perspectives. Finally, Section \ref{sec:methods} concludes the paper with pertinent methodological comments.

All data and examples from this paper are available at: \url{https://orgdlabs.com/neuroQL}.

\section{Results} \label{sec:results}

\subsection{NeuroQL by Example} \label{sec:neuroql_by_example}

You can load the NeuroQL language inside a python environment as the one we provide for our readers at: \url{https://orgdlabs.com/neuroql_eg} by running the following import statement (shown in line 1 of listing \ref{lst:import}):

\lstset{style=mystyle2}
\begin{lstlisting}[label={lst:import}, caption=Importing NeuroQL into Python and loading product properties and reviews]
from NeuroQL import *
NeuroQL.load('asin_key_properties.csv','asin_reviews.csv')
\end{lstlisting}

Then as shown in line 2 of listing \ref{lst:import} you can load data (\eg from csv files) to populate your knowledge base. In this case we load product properties and reviews in line 2, which have the following general format:

\lstset{style=datastyle}
\begin{lstlisting}[label={lst:props_eg}, caption=Product properties excerpt]
  ...
B00001P4ZH,title,koss portapro headphones with case
B00001P4ZH,price,39.36
B00001P4ZH,brand,koss
B00001P4ZH,stars,4.7
B00001P4ZH,total_reviews,14549
B00001P4ZH,item_weight,2.79 
B00001P4ZH,item_weight_units,ounces
B00001P4ZH,item_model_number,6303157
B00001P4ZH,is_discontinued_by_manufacturer,no
  ...
\end{lstlisting}

This data consists of a product identifier (like \textit{B00001P4ZH}), which is called an \textit{asin} (Amazon Standard Identification Number) in the \url{amazon.com} product catalog \cite{bausch2003amazon}. The identifier is followed by the name of a property (such as \textit{title}, or \textit{price}) together with that property's value (\eg \textit{39.36} in the case of the \textit{price} property on line 3 above). NeuroQL is able to load arbitrarily nested n-tuples of any length, but for the sake of simplicity we use here a flat representation of the form: \textit{object.property = value} familiar to programmers from OO paradigms as well as engineers working with triple-based knowledge systems \textit{(subject,predicate,object)} \cite{berners2008n3logic}. If we need to load new items explicitly (instead from a pre-saved format) we can invoke the NeuroQL \textit{\lstinline[style=datastyle]{fact}} primitive as follows: \textit{\lstinline[style=datastyle]{fact(('B00001P4ZH','price',39.36))}}. Alternatively we can declare the id: \textit{\lstinline[style=datastyle]{B00001P4ZH = ids()}} and simply state \textit{\lstinline[style=datastyle]{B00001P4ZH.price == 39.36}}. 

After having loaded the data, we can perform our first simple query using NeuroQL as follows:

\lstset{style=datastyle}
\begin{lstlisting}[label={lst:search_id}, caption=Retrieving all available product properties by id]
B00001P4ZH = ids()
_property, _value = vars()
search(B00001P4ZH._property == _value)
\end{lstlisting}

Here in listing \ref{lst:search_id} we define an id of interest on line 1 and declare our variables of interest on line 2. In this case we would like to retrieve all available properties and their values regarding a particular id. To do so, on line 3 we invoke the \textit{search} NeuroQL primitive, which takes as input a NeuroQL expression, and returns all possible combinations of values from our knowledge base that satisfy our input expression.  

Thus our search results always consist of a list of mappings (such as dictionaries) that bind our unknown variables (seen on listing \ref{lst:search_id_results}) to the values that satisfy our expression:

\lstset{style=datastyle}
\begin{lstlisting}[label={lst:search_id_results}, caption=Product properties retrieved by id]
{'?property': 'stars', '?value': 4.7},
{'?property': 'item_weight', '?value': 2.79},
{'?property': 'total_reviews', '?value': 14549},
{'?property': 'title', '?value': 'koss portapro ..'},
{'?property': 'price', '?value': 39.36},
{'?property': 'review', '?value': '882b1e2745a47...'},
{'?property': 'brand', '?value': 'koss'},
{'?property': 'is_discontinued_by_manufacturer','?value': 'no'},
{'?property': 'item_model_number', '?value': 6303157},
{'?property': 'review', '?value': 'ce76793f036494...'},
{'?property': 'item_weight_units', '?value': 'ounces'},
{'?property': 'review', '?value': 'd040f2713caa2...'}
\end{lstlisting}

This type of binding is achieved through symbolic unification \cite{robinson1965machine,robinson2001handbook}, which we will be extending later on with a set of neuronal primitives. 

Continuing with our examples, NeuroQL also allows for the creation of more complex queries from simpler ones, with additional means of combination as shown in listing \ref{lst:match_title_price}:

\lstset{style=mystyle2}
\begin{lstlisting}[label={lst:match_title_price}, caption=Searching for headphones with a price around 30\$] 
_asin,_total_reviews,_price,_title,
_review,_review_text,_answers=vars()

search(
    bm25_match(_asin.title==_title,'headphones',80),
    _asin.price==_price, 
    op_filter(lambda e: abs(e['?price'] - 30) < 10),
)
\end{lstlisting}

Here on lines 1 and 2 we define a set of variables that we want to find the bindings of (for this and subsequent examples). Then on lines 4 through 8 we are asking for the ids of headphones whose price is around 30 dollars. To express this query in NeuroQL we form a conjunction of expressions (by passing each expression as a separate argument to our \textit{search} primitive). The first such expression performs a \textit{bm25} \cite{robertson2009probabilistic} matching against all product titles, returning the top 80 results (line 5).
Similarly, on line 6 we are asking for the \textit{\_price} variable to be bound to the price of the products that we have found thus far. Finally on line 7, we are filtering our results to only include products whose price is within 10 dollars of our target price.  

Notice here, that the id (\ie the \textit{\_asin}), the product \textit{\_title} as well as the product \textit{\_price}, are all variables. These are the variables which we are asking NeuroQL to bound for us. Each binding will correspond to specific values from our knowledge base that jointly satisfy our expressions, getting us the following results (listing \ref{lst:match_title_price_results}): 

\lstset{style=datastyle}
\begin{lstlisting}[label={lst:match_title_price_results}, caption=Output example for a simple query (showing 2 sample results out of a total of 10)]
...
{'?asin': 'B00001P4ZH',
'?title': 'koss portapro headphones with case',
'?price': 39.36},
...
{'?asin': 'B000AJIF4E',
'?title': 'sony mdr7506 professional large diaphragm headphone',
'?price': 29.99}
...
\end{lstlisting}

Building towards our complete example from Figures \ref{fig:question} and \ref{fig:translation} we can now ask (listing \ref{lst:inc_reviews_manufacturing}) for \textit{headphones at around 30 dollars, having minimum 14K reviews, that are not discontinued by the manufacturer}:

\lstset{style=mystyle2}
\begin{lstlisting}[label={lst:inc_reviews_manufacturing}, caption=Refining our search to include total number of reviews and manifacturing details] 
search(
    bm25_match(_asin.title==_title,'headphones',80),
    _asin.price==_price, 
    op_filter(lambda e: abs(e['?price'] - 30) < 10),
    _asin.total_reviews==_total_reviews,  
    op_filter(lambda e: e['?total_reviews'] >= 14000),
    _asin.is_discontinued_by_manufacturer=='no',
)
\end{lstlisting}

This is achieved by extending our previous query (of listing \ref{lst:match_title_price}) with lines 5 through 7 on listing \ref{lst:inc_reviews_manufacturing}. Describing a \textit{\_total\_reviews} variable to be bound (for our already retrieved products) on line 5. Filtering the results further depending on the total number of reviews on line 6. Then finally, on line 7 adding the additional constraint that the products we are looking for should not have been discontinued. The results from this extended query are seen in listing \ref{lst:results_inc_reviews_manufacturing}:

\lstset{style=datastyle}
\begin{lstlisting}[label={lst:results_inc_reviews_manufacturing}, caption=Output example for a refined query (showing all 3 results)]
{'?asin': 'B00001P4ZH',
'?title': 'koss portapro headphones with case',
'?price': 39.36,
'?total_reviews': 14549},
{'?asin': 'B0007XJSQC',
'?title': 'sennheiser hd201 lightweight over-ear .. headphones',
'?price': 24.95,
'?total_reviews': 14980},
{'?asin': 'B000AJIF4E',
'?title': 'sony mdr7506 professional large diaphragm headphone',
'?price': 29.99,
'?total_reviews': 22071}
\end{lstlisting}

\subsubsection{Neuro-Symbolic Composition}\label{sec:neuro-symbolic-composition}

Up to this point we have been seeing how NeuroQL can handle the objective components of natural language questions. Now we will extend our previous example (listing \ref{lst:inc_reviews_manufacturing}) to handle the entirety of an inter-subjective question as the one we described in Figures \ref{fig:question} and \ref{fig:translation}. Namely, \textit{``How is the bass for headphones at around 30 dollars having minimum 14K reviews that is not discontinued?''}. To do so, we start by extending our previous example on line 8 of listing \ref{lst:full-eg} where we bind a \textit{\_review} variable to the reviews of the products we have found so far. 

Then on line 9 we invoke the neural primitive \lstinline[style=mystyle2]{neural_match}, taking the following algorithmic steps to extend classical unification: 
\begin{enumerate}
\item \textit{\lstinline[style=mystyle2]{neural_match} will receive the sub-query \lstinline[style=mystyle2]{_review.text==_review_text} as its first argument. It will use this sub-query to create a set of \textit{(id, document)} pairs (in this case \lstinline[style=mystyle2]{_review,_review_text} pairs) binding the variable \lstinline[style=mystyle2]{_review_text} in the process.}
\item \textit{It will then try to match the bindings of \lstinline[style=mystyle2]{_review_text} against the subjective component of the initial query, in this case: \lstinline[style=mystyle2]{'how is the bass?'}, which is the second argument we passed to our \lstinline[style=mystyle2]{neural_match} primitive. With default settings this match will be performed by a DPR (Dense Passage Retriever) \cite{karpukhin2020dense}, using question \& context encoders trained with the Natural Questions dataset \cite{lee-etal-2019-latent,kwiatkowski-etal-2019-natural}.} 
\item \textit{Finally, \lstinline[style=mystyle2]{neural_match} will filter all query bindings thus far (up to line 11 of listing \ref{lst:full-eg}) to include only the top 5 results or our DPR \lstinline[style=mystyle2]{_review_text} match (using the third argument on line 10).}
\end{enumerate}%

\lstset{style=mystyle2}
\begin{lstlisting}[label={lst:full-eg}, caption=Creating an inter-subjective query by filtering and scanning reviews to answer a particular question] 
search(
    bm25_match(_asin.title==_title,'headphones',80),
    _asin.price==_price, 
    op_filter(lambda e: abs(e['?price'] - 30) < 10),
    _asin.total_reviews==_total_reviews,  
    op_filter(lambda e: e['?total_reviews'] >= 14000),
    _asin.is_discontinued_by_manufacturer=='no',
    _asin.review==_review, 
    neural_match(
      _review.text==_review_text,'how is the bass?',5
    ),
    neural_extract(
      _answers, _review.text==_review_text,'how is the bass?',2
    )
)
\end{lstlisting}

Thus far we have the top 5 reviews that match our question for the exact subset of products satisfying our objective constraints (up to line 11). We can now proceed to extract relevant opinions for our inter-subjective question using the \lstinline[style=mystyle2]{neural_extract} primitive (lines 12 to 14 of listing \ref{lst:full-eg}), as follows: 

\begin{enumerate}
  \item \textit{\lstinline[style=mystyle2]{neural_extract} will first receive the name of a new variable to bind (in this case the variable \lstinline[style=mystyle2]{_answers}) with the extracted text that matches its targeted question.}
  \item \textit{It will then use the sub-query \lstinline[style=mystyle2]{_review.text==_review_text} passed as a second argument, to create a new set of \textit{(ids, documents)} matching the sub-query (as we did before).}
  \item \textit{Then with the third argument (the subjective sub-component \lstinline[style=mystyle2]{'how is the bass?'}) it will try to extract the answer to this question from the \lstinline[style=mystyle2]{_review_text} documents. It will do so using a Reader model, such as MiniLM \cite{wang2020minilm} initially trained on the SQuAD 2.0 dataset \cite{rajpurkar2018know} and further fine-tuned on the NeuroQL training set to improve its performance (as we describe in Section \ref{sec:experiments_and_validation}).}
  \item \textit{Finally, \lstinline[style=mystyle2]{neural_extract} will filter all query bindings to include only the top 2 results extracted by the Reader (using the forth argument on line 13).}
\end{enumerate}

The final results of our inter-subjective query are seen in listing \ref{lst:full-eg-results} with a sample of the final variable bindings. These now include both objective (\textit{title, price, total\_reviews, manufacturing}) and subjective (\textit{review, answers}) components that jointly satisfied our query's constraints. As an example the most pertinent opinion for product \lstinline[style=datastyle]{B000AJIF4E} (satisfying our constraints regarding \textit{title, price, total\_reviews, manufacturing} \etc)  is that the \lstinline[style=datastyle]{'Bass is amazing'}, while for product \lstinline[style=datastyle]{B00001P4ZH} is that the \lstinline[style=datastyle]{'Bass is weak as expected'}.

\lstset{style=datastyle}
\begin{lstlisting}[label={lst:full-eg-results}, caption=Output example for an inter-subjective query]
{'?asin': 'B000AJIF4E',
'?title': 'sony mdr7506 professional large diaphragm headphone',
'?price': 29.99,
'?total_reviews': 22071,
'?review': '5e96b0052898fe667cf622888fc5af69',
...
'?answers': {'answer': 'Bass is amazing',...}}
...
{'?asin': 'B00001P4ZH',
'?title': 'koss portapro headphones with case',
'?price': 39.36,
'?total_reviews': 14549,
'?review': 'd040f2713caa2aff0ce95affb40e12c2',
...
'?answers': {'answer': 'Bass is weak as expected',...}}
\end{lstlisting}

We complete our examples by showing how to define and use inference rules in NeuroQL \footnote{A formal model of the syntax and semantics of inference rules in NeuroQL will be part of a follow up paper.}. Lines 1 through 6 of listing \ref{lst:rull-eg} define an inference rule using the primitive \lstinline[style=mystyle2]{rule}. The first argument on line 1 is the sub-query (\lstinline[style=mystyle2]{_asin.well_ranked == True}) that defines the conclusion that the rule can infer \textit{\textbf{if}} the rest of the arguments are satisfied. The rest of the arguments being sub-queries themselves forming a conjunction.

Whenever the head of this rule unifies with a sub-query that we are currently looking to bind (like the sub-query on line 11 of our \lstinline[style=mystyle2]{search} primitive) the rule will be tested. This means that there will be a nested search that will try to satisfy the rule's conjunction given the current bindings and return new bindings if needed. 

\lstset{style=mystyle2}
\begin{lstlisting}[label={lst:rull-eg}, caption=Incorporating a rule to infer well ranked products during search] 
rule(_asin.well_ranked == True,
  _asin.total_reviews==_total_reviews,
  op_filter(lambda e: e['?total_reviews'] >= 20000),
  _asin.stars==_stars,
  op_filter(lambda e: e['?stars'] >= 4.0)
)
search(
    bm25_match(_asin.title==_title,'headphones',80),
    _asin.price==_price, 
    op_filter(lambda e: abs(e['?price'] - 30) < 10),
    _asin.well_ranked == True,
    _asin.is_discontinued_by_manufacturer=='no',
    _asin.review==_review, 
    neural_match(
      _review.text==_review_text,'how is the bass?',5
    ),
    neural_extract(
      _answers, _review.text==_review_text,'how is the bass?',2
    )
)
\end{lstlisting}

In plain english this rule states that a product is well ranked if it has at least 20K reviews and a rating of at least 4.0 stars. By using this rule on line 11 of listing \ref{lst:rull-eg} instead of our previous less strict criteria (on lines 5 and 6 of listing \ref{lst:full-eg}) we get a more constraint result (seen on listing \ref{lst:rule-eg-results}).

\lstset{style=datastyle}
\begin{lstlisting}[label={lst:rule-eg-results}, caption=Inter-subjective query results using rules for well ranked products during inference]
...
{'?asin': 'B000AJIF4E',
'?title': 'sony mdr7506 professional large diaphragm headphone',
'?price': 29.99,
'?total_reviews': 22071,
'?review': '5e96b0052898fe667cf622888fc5af69',
...
'?answers': {'answer': 'Bass is amazing',...}}
\end{lstlisting}

\subsection{The NeuroQL Architecture} \label{sec:neuroql_architecture}

\begin{figure}[h!] 
  \centering
  \advance\leftskip-2.9cm 
  \includegraphics[width=1.4\textwidth]{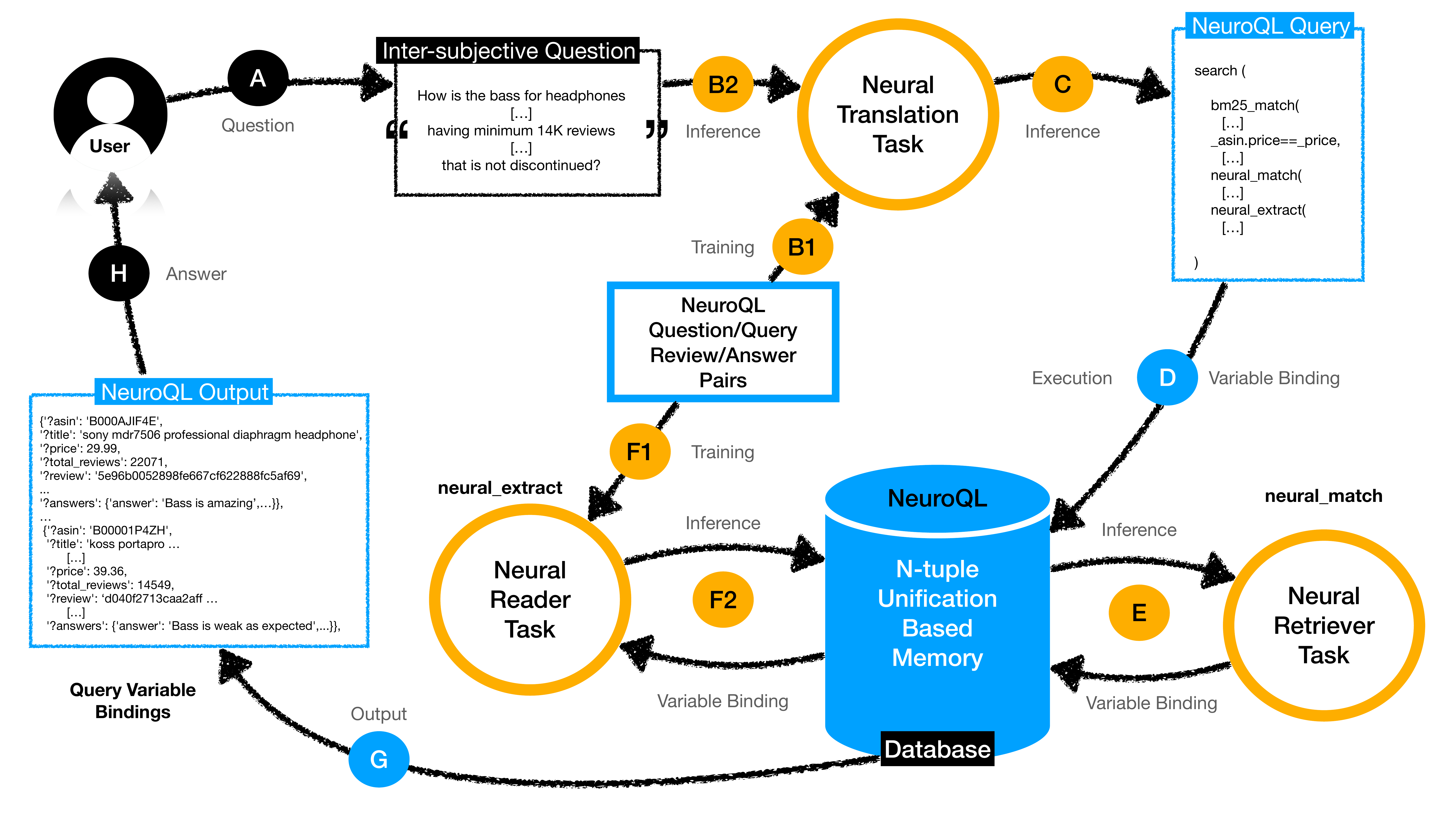}
  \caption{NeuroQL Architecture: Starting from an inter-subjective question the NeuroQL architecture firsts infers the equivalent NeuroQL query using a neural translation network that has been fine-tuned for this task. Subsequently the query is executed against the NeuroQL database, that employs both symbolic (through unification) and neural reasoning (through a retriever and a reader network) to incrementally bind the query variables and return an answer to the user.}
  \label{fig:neuroql-architecture}
\end{figure}

Our two main hypotheses (as detailed in Section \ref{sec:intro}) are that (\textbf{\textit{H1}}) it is possible to automatically translate inter-subjective questions from natural language into NeuroQL and (\textbf{\textit{H2}}) that NeuroQL can concisely extend unification with neural reasoning to produce satisfactory answers for inter-subjective questions. To test (\textbf{\textit{H1}}) and (\textbf{\textit{H2}}) we devised the following architecture (seen in Figure \ref{fig:neuroql-architecture}) that serves as a baseline solution for our inter-subjective Q\&A task. Starting at the top left corner of Figure \ref{fig:neuroql-architecture} we see a user submitting an inter-subjective question in natural language \circled[black]{A}. This question during inference \circled[orange]{B2} will be passed to a translation model (in our case a fine-tuned CodeT5 model \cite{wang2021codet5}) that will attempt to translate the question into a NeuroQL query \circled[orange]{C}. The translation model is trained \circled[orange]{B1} for this downstream task using the NeuroQL question/query dataset (detailed in Sections \ref{sec:neuroql_dataset} and \ref{sec:experiments_and_validation}). Subsequently, our architecture will attempt to execute this query \circled[blue]{D} which (as we saw in Section \ref{sec:neuroql_by_example}) means finding all possible bindings within the NeuroQL database that satisfies the query's main conjunction. The NeuroQL database itself is an in-memory n-tuple store upon which our extended unification algorithm is applied. These extensions include the \lstinline[style=mystyle2]{neural_match} \circled[orange]{E} and \lstinline[style=mystyle2]{neural_extract} \circled[orange]{F2} primitives. These primitives are based on a retriever model (in our case a DPR \cite{karpukhin2020dense} trained with the Natural Questions dataset \cite{lee-etal-2019-latent,kwiatkowski-etal-2019-natural}) and a reader model (a MiniLM \cite{wang2020minilm} initially trained on the SQuAD 2.0 dataset \cite{rajpurkar2018know}). Using these models our unification engine will attempt to further bind and filter the neuronal sub-queries of a logical conjunction (as we saw in Section \ref{sec:neuro-symbolic-composition}). During training our reader has been further fine-tuned \circled[orange]{F1} on the NeuroQL training set (using our review/answer pairs) to improve its performance (as we further describe in Sections \ref{sec:neuroql_dataset} and \ref{sec:experiments_and_validation}). Finally, when all possible bindings have been found \circled[blue]{G} an answer \circled[black]{H} will be returned to our user.

Listings \ref{lst:translate} and \ref{lst:dynEval} show step by step how we can first translate a natural language question into NeuroQL (lines 1 to 3 of listing \ref{lst:translate}) invoking our primitive \lstinline[style=mystyle2]{NeuroQL.translate}. Then, how to dynamically evaluate the resulting query (in line 1 of listing \ref{lst:dynEval}) to get our results (using \lstinline[style=mystyle2]{NeuroQL.eval}). This is done with both cases matching our previous results from listings \ref{lst:full-eg} and \ref{lst:full-eg-results}. Finally, listing \ref{lst:answer} show us how both translation and evaluation can be expressed in a single step (line 1 of listing \ref{lst:answer}) by simply invoking our primitive \lstinline[style=mystyle2]{NeuroQL.answer}.
\vspace{0.5em}

\lstset{style=mystyle2}
\begin{lstlisting}[label={lst:translate}, caption=Translating an inter-subjective question into a NeuroQL query] 
  query = NeuroQL.translate(
    'How is the bass for headphones at around [...] ?'
  )
  print(query)

  Output: #########################################
  search(
    bm25_match(_asin.title==_title,'headphones',80),
    _asin.price == _price, 
    op_filter(lambda e: abs(e['?price'] - 30) < 10),
    _asin.total_reviews == _total_reviews, 
    op_filter(lambda e: e['?total_reviews'] >= 14000),
    _asin.is_discontinued_by_manufacturer=='no',
    _asin.review == _review, 
    neural_match(
      _review.text == _review_text,'how is the bass?',5
    ),
    neural_extract(
      _answers, _review.text==_review_text,'how is the bass?',2
    )
  )
  ###################################################
\end{lstlisting} 

\lstset{style=datastyle}
\begin{lstlisting}[label={lst:dynEval}, caption=Dynamically evaluating a generated NeuroQL query]] 
  NeuroQL.eval(query)

  Output: #########################################
{'?asin': 'B000AJIF4E',
'?title': 'sony mdr7506 professional large diaphragm headphone',
'?price': 29.99,
'?total_reviews': 22071,
'?review': '5e96b0052898fe667cf622888fc5af69',
...
'?answers': {'answer': 'Bass is amazing',...}}
...
{'?asin': 'B00001P4ZH',
'?title': 'koss portapro headphones with case',
'?price': 39.36,
'?total_reviews': 14549,
'?review': 'd040f2713caa2aff0ce95affb40e12c2',
...
'?answers': {'answer': 'Bass is weak as expected',...}}
  ###################################################
\end{lstlisting}

\lstset{style=datastyle}
\begin{lstlisting}[label={lst:answer}, caption=Combining translation and evaluation of inter-subjective questions in a single call] 
  NeuroQL.answer(
    'How is the bass for headphones at around [...] ?'
  )
  
  Output: #########################################
{'?asin': 'B000AJIF4E',
'?title': 'sony mdr7506 professional large diaphragm headphone',
'?price': 29.99,
'?total_reviews': 22071,
'?review': '5e96b0052898fe667cf622888fc5af69',
...
'?answers': {'answer': 'Bass is amazing',...}}
...
{'?asin': 'B00001P4ZH',
'?title': 'koss portapro headphones with case',
'?price': 39.36,
'?total_reviews': 14549,
'?review': 'd040f2713caa2aff0ce95affb40e12c2',
...
'?answers': {'answer': 'Bass is weak as expected',...}}
  ###################################################
\end{lstlisting}
\vspace{-1em}
\subsection{The NeuroQL Dataset}  \label{sec:neuroql_dataset} The NeuroQL dataset extends previous work on Q\&A systems focused on metadata \cite{ni2019justifying} and subjective \cite{bjerva2020subjqa} information to include \textit{1505} inter-subjective questions and their translation into neuro-symbolic queries (\eg in listing \ref{lst:qq-eg}). These \textit{(question, query)} pairs are coupled with a detailed knowledge base of \textit{4250} properties (\eg in listing \ref{lst:props_eg}) for \textit{500} different products, including \textit{1583} reviews and \textit{1627} ground truths for Q\&A extraction. 

\lstset{style=datastyle}
\begin{lstlisting}[label={lst:qq-eg}, caption={A \textit{(question, query)} pair from the NeuroQL dataset}]
B00001P4ZH,question,0514ee34 ...
0514ee34 ..., text,For headphones model number 6303157 ...
0514ee34 ..., query,"search(
  bm25_match( ...
  ...
  neural_match( ...
  neural_extract( ... 
)"
\end{lstlisting} 

\begin{figure}[h!]%
  \centering
  \advance\leftskip-3.0cm  
  \advance\rightskip-3.0cm
  \subfloat[\centering Number Questions per Objective Property]{{\includegraphics[width=9cm]{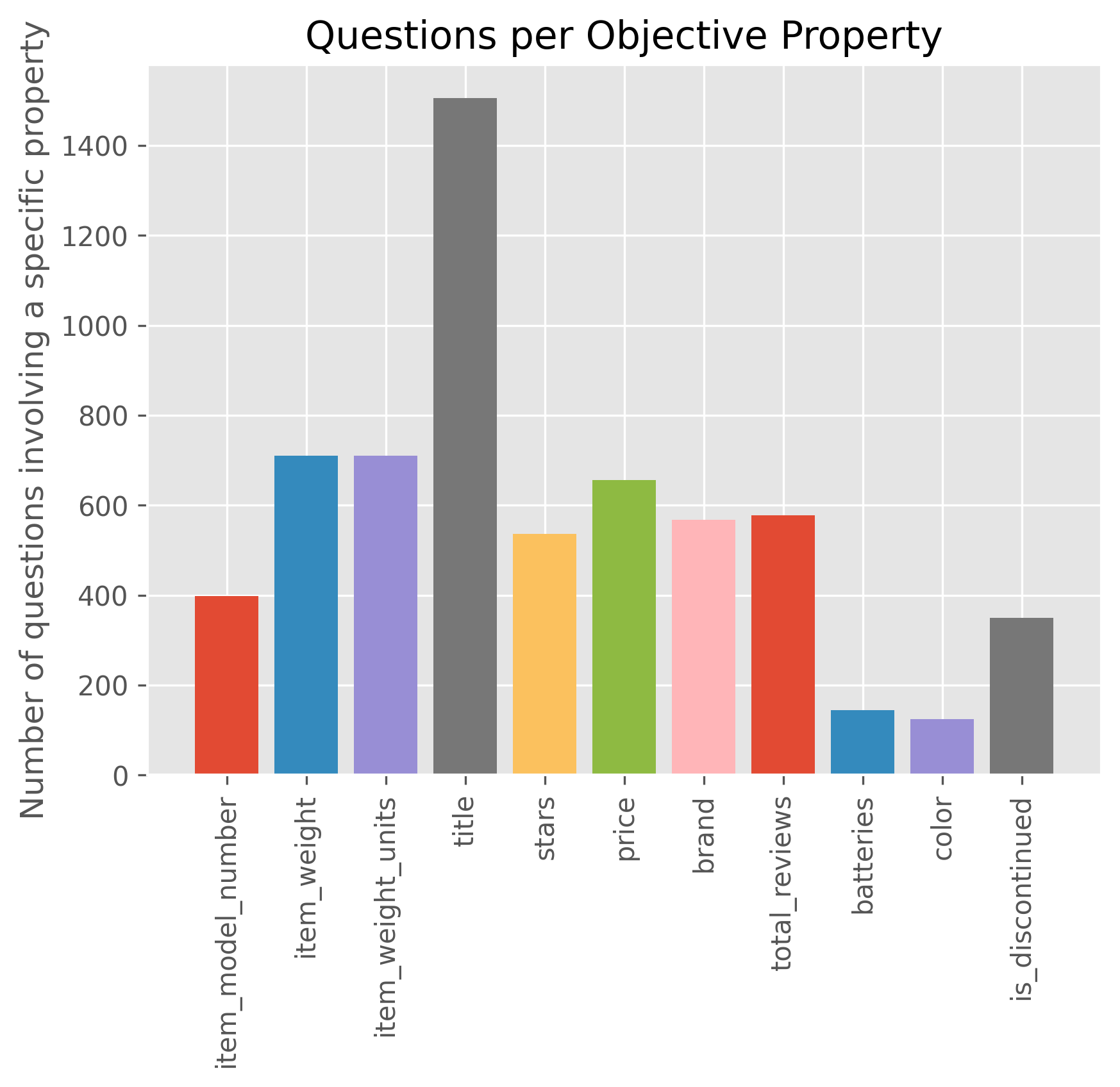} }}%
  \qquad
  \subfloat[\centering Subjective Component Statistics of Inter-Subjective Questions]{{\includegraphics[width=9cm]{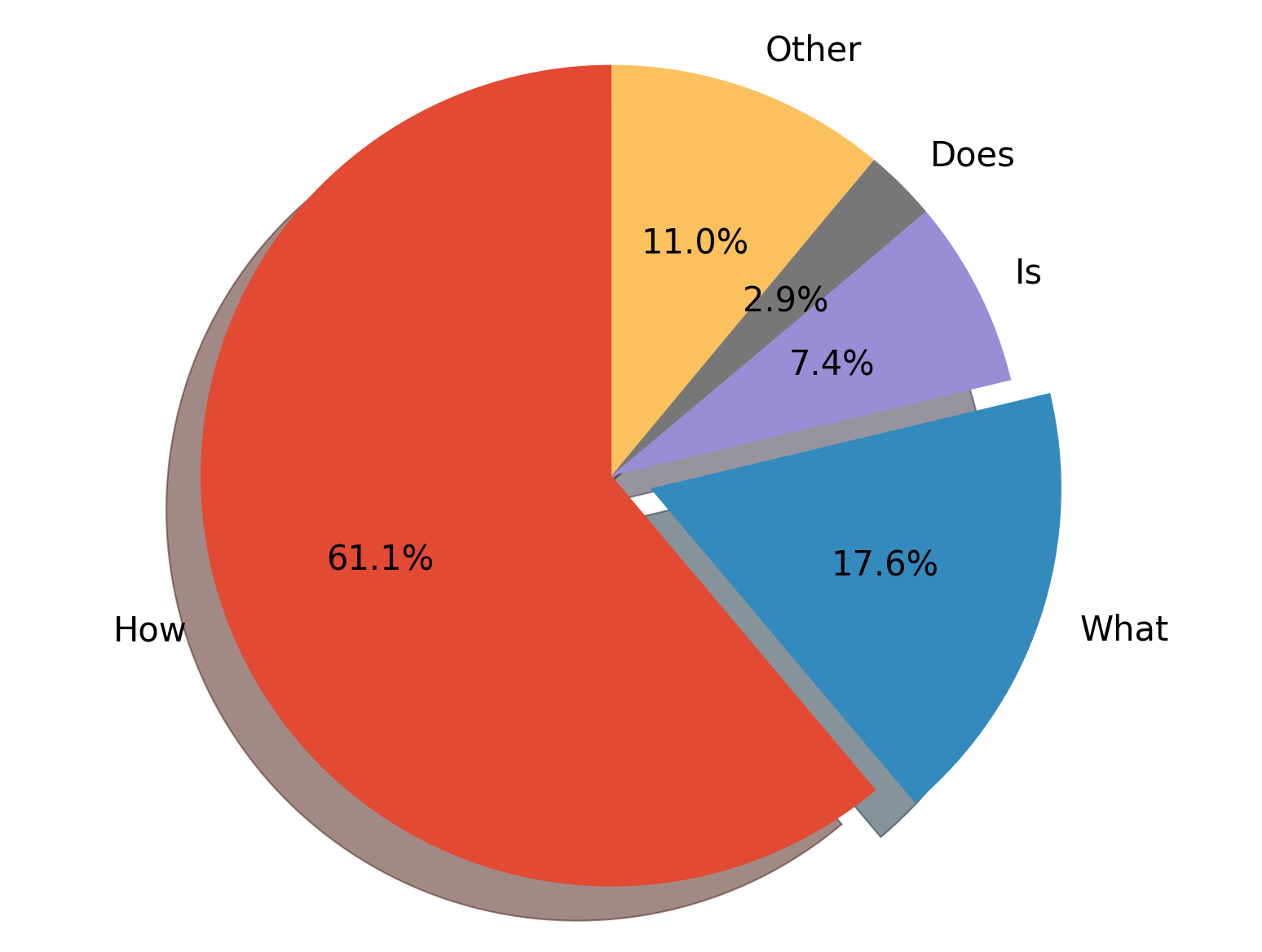} }}%
  \caption{Breakdown of properties and subjective components in the NeuroQL Dataset}%
  \label{fig:ds-breakdown}%
\end{figure}

On Table \ref{obj-properties-table} we list all the categories of objective properties that we included in our dataset with a short description for each. Each category is followed by either its domain or a sample value for reference. Finally on Figure \ref{fig:ds-breakdown} we present a breakdown of our dataset in terms of questions involving a specific objective property (on the left of Figure \ref{fig:ds-breakdown}). On the right (similarly to \cite{tunstall2022natural}) we give statistics regarding the first word of the subjective component present in our inter-subjective questions.

\begin{table}
  \centering
  \begin{tabular}{lll}  
    \toprule
    \multicolumn{2}{c}{Product Properties}                   \\
    \cmidrule(r){1-2}
    Name     & Description     & Sample Values\\
    \midrule
    title & The product's title as reported online  & \textit{\eg apple magic mouse}    \\
    brand & The product's brand  & \textit{\eg audio-technica}    \\
    item\_model\_number & Manufacturer's Serial No or other id  & \textit{\eg 6229a003aa}    \\
    price  & The product's price as reported online & \textit{in range} $\mathit{[1.95, 999.0]}$    \\
    stars   & The product's average rating & \textit{in range} $\mathit{[2.8, 5]}$  \\
    total\_reviews  & Total number of reviews of a product & \textit{in range} $\mathit{[2, 134717]}$    \\
    color & The product's color as reported online  & \textit{\eg amethyst gray}    \\
    item\_weight    & The product's price as reported online & \textit{in range} $\mathit{[0.01, 73.2]}$  \\
    item\_weight\_units & Units of item\_weight & $\in \{'kilograms','ounces', 'pounds'\}$  \\
    batteries & Number of batteries included  & \textit{in range} $\mathit{[1, 3]}$    \\
    is\_discontinued     & Continued Manufacturing & $\in \{'no','yes'\}$  \\
    \bottomrule
  \end{tabular}
  \vspace{0.8em}
  \caption{NeuroQL Dataset: Type, Description and Sample Values of Objective Properties}
  \label{obj-properties-table}
\end{table}

\subsection{Experiments and Validation} \label{sec:experiments_and_validation}

There are two distinct experimental tasks that the NeuroQL dataset makes possible. The first is the \textit{NeuroQL Translation Task} where the goal is to translate inter-subjective questions posed in natural language into the neuro-symbolic code that can answer them. The task takes into consideration \textit{(question, query)} pairs and can be evaluated using metrics such as \textit{sacreBleu} \cite{post2018call} and four-gram or tri-gram precision, common in machine translation tasks \cite{tunstall2022natural}.

The second task is the \textit{NeuroQL Query Task} where the goal is to fine-tune neural primitives such as the \lstinline[style=mystyle2]{neural_extract} sub-query to produce as many accurate results as possible. The task takes into consideration \textit{(query, answers)} pairs and can be evaluated using metrics such as \textit{recall}, \textit{em} and \textit{f1} common in Q\&A extraction tasks \cite{tunstall2022natural}.

\begin{figure}[h!]%
  \centering
  \advance\leftskip-3.0cm 
  \advance\rightskip-3.0cm
  \subfloat[\centering SacreBleu score (training, validation, test sets) for translating inter-subjective questions posed in natural language into code for NeuroQL.]{{\includegraphics[width=9cm]{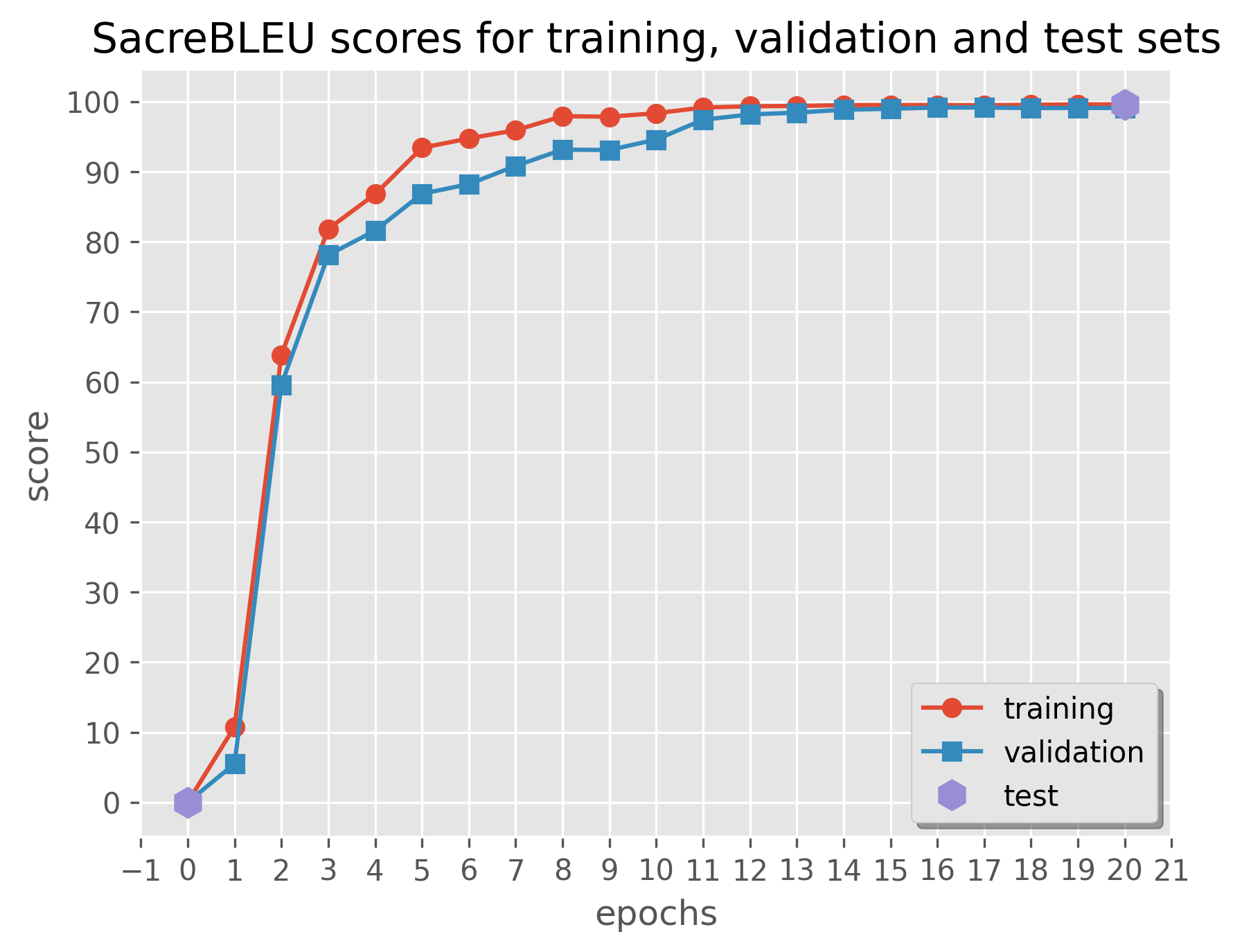} }}%
  \qquad
  \subfloat[\centering Four-gram and tri-gram precision per epoch for our validation set, when translating inter-subjective questions into NeuroQL. On the upper right corner we also report the test set four-gram precision after training.]{{\includegraphics[width=9cm]{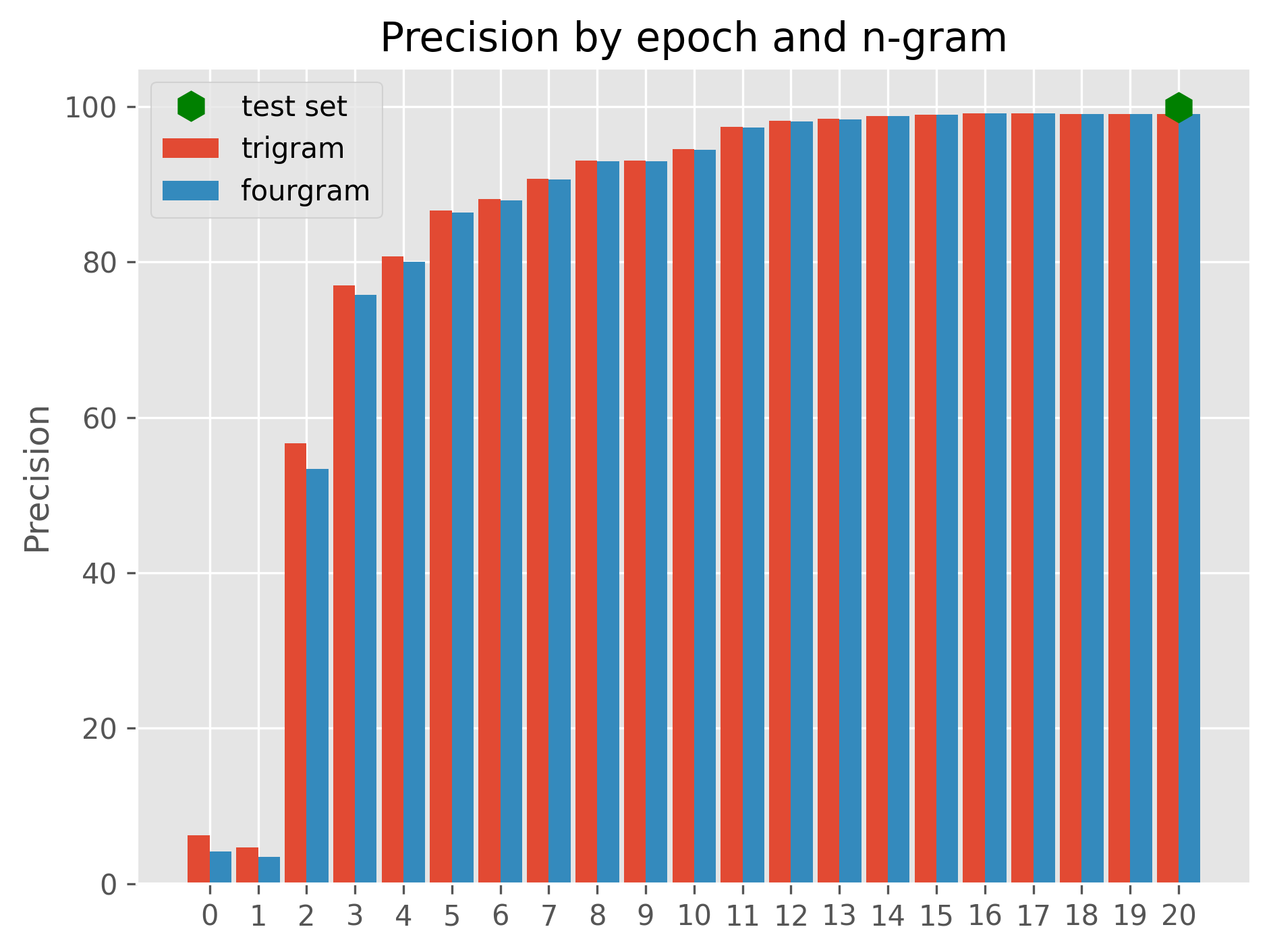} }}%
  \captionsetup{justification=centering}
  \caption{Our Experimental Results for the NeuroQL Translation Task including sacreBleu, four-gram and tri-gram precision metrics.}%
  \label{fig:translation-experiments}%
\end{figure}

For our \textit{NeuroQL Translation Task} we fine-tune a CodeT5 model \cite{wang2021codet5} for \textit{20 epochs} aiming to translate inter-subjective questions into NeuroQL code.  We use a maximum input/output length of \textit{512} tokens that fits our \textit{(question,query)} pairs, a batch size of \textit{64} and an \textit{1e-4} learning rate with a \textit{linear} scheduler and an \textit{AdamW} optimizer \cite{loshchilov2017decoupled}. As we can see in the the left part of Figure \ref{fig:translation-experiments}, the model is significantly improving up until epoch 11 (without over-fitting), after which point the returns are diminishing leading to a near perfect sacreBleu score at epoch 20 for both training and validation sets. The test set evaluation after epoch 20 confirms our model's performance. We split our dataset between training \textit{(80\%)}, validation \textit{(10\%)} and test \textit{(10\%)} sets using unique product ids for each slice to ensure no leakage. The four-gram and tri-gram precisions reported on the right side of Figure \ref{fig:translation-experiments} follow our sacreBleu observations as expected. Given these results our initial hypothesis (\textbf{\textit{H1}}) regarding the feasibility of translating inter-subjective questions into neuro-symbolic queries has been validated.

\begin{figure}[h!]%
  \centering
  \advance\leftskip-3.0cm 
  \advance\rightskip-3.0cm
  \subfloat[\centering Recall scores for dense retriever (\ie neural\_match) for different values of top\_k scored results.]{{\includegraphics[width=9cm]{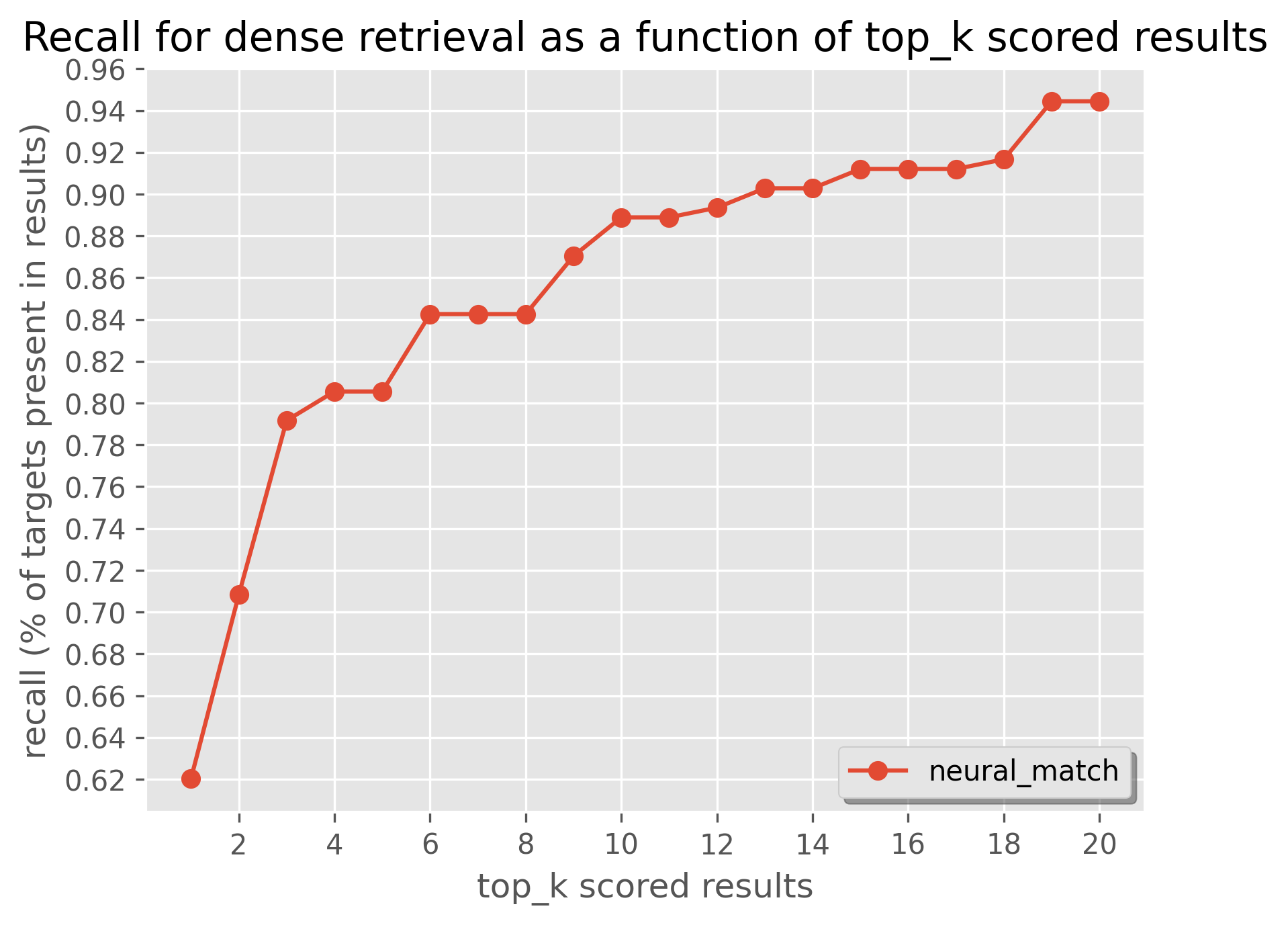} }}%
  \qquad
  \subfloat[\centering EM \& F1 scores for reader (\ie neural\_extract) for different values of top\_k scored results.]{{\includegraphics[width=9cm]{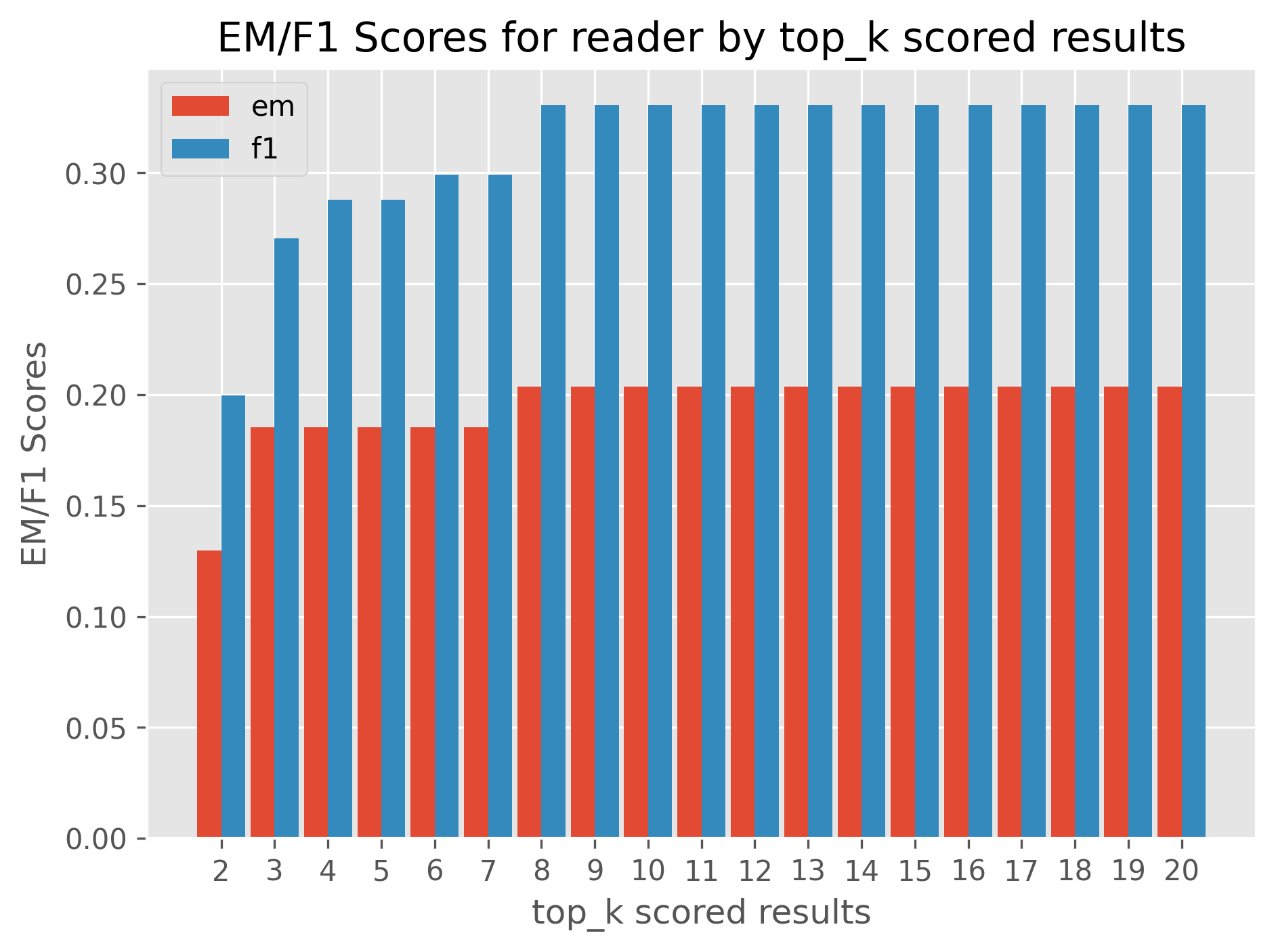} }}%
  \captionsetup{justification=centering}
  \caption{Our Experimental Results for the NeuroQL Query Task including recall, em and f1 scores for our DPR and Reader models.}%
  \label{fig:dpr-reader-experiments}%
\end{figure}

For our \textit{NeuroQL Query Task}, we first measure the impact our DPR \cite{karpukhin2020dense} model trained with the Natural Questions dataset \cite{lee-etal-2019-latent,kwiatkowski-etal-2019-natural}) has on our test set, for different values of \textit{top\_k} results returned by the retriever.
We then fine-tune our reader model (a MiniLM \cite{wang2020minilm} initially trained on the SQuAD 2.0 dataset \cite{rajpurkar2018know}) to improve its performance for different values of \textit{top\_k} results returned by the reader. On the left part of Figure \ref{fig:dpr-reader-experiments} we can see that our DPR model performs quite well at \textit{70\%} recall even in the case of only the \textit{top\_2} results returned. Yet, it does not surpass the \textit{90\%} mark before \textit{top\_13} results are returned, reaching a \textit{94\%} recall score at \textit{top\_20}. Using this last recall score from our retriever, we then evaluate our fine-tuned reader model, observing as expected an increase at \textit{em} and \textit{f1} scores as \textit{top\_k} increases. Yet, the best values observed were \textit{0.20} for \textit{em} and \textit{0.33} for \textit{f1}. These were both reached at \textit{top\_8} results returned, being comparable to scores for extractive Q\&A on user reviews reported by \cite{tunstall2022natural}, using the same model. 

Prior to this evaluation we fine-tuned our MiniLM \cite{wang2020minilm} reader model for \textit{3 epochs} in order to improve its extraction accuracy. We used a \textit{384}-token sequence length with a \textit{128}-token document stride and a batch size of \textit{16}, including the ability to return no answers when predicting results. We set the learning rate at \textit{1e-5} with a \textit{0.2} warmup, using the same dataset split as before. Given the above results our initial hypothesis (\textbf{\textit{H2}}) regarding the feasibility of extending unification with neural primitives for answering inter-subjective questions has only been partially validated. Neuro-symbolic unification as presented in this work, can indeed be used to execute inter-subjective neuro-symbolic queries. Yet these results present only a first baseline. There is more work needed to improve the accuracy of extracted answers, particularly regarding the \lstinline[style=mystyle2]{neural_extract} primitive.

\subsection{Related Work} \label{sec:related_work}

There is a significant body of work dedicated to translating natural language questions to queries for established symbolic languages such as sql \cite{kim2020natural,brunner2021valuenet,xuan2021sead,wu2021data} and sparql \cite{yin2021neural,saparina2021sparqling}. These solutions have only one neuronal component for translation, going from: \textit{neural translation $\vartriangleright$  symbolic code $\vartriangleright $  symbolic reasoning}. Our contribution takes a step further by translating natural language into neuro-symbolic code, closing the loop between symbolic and neuronal reasoning: \textit{neural translation $\vartriangleright $  neuro-symbolic code $\vartriangleright $ neuro-symbolic reasoning}. 

Deep-prolog \cite{manhaeve2018deepproblog} and TensorLog \cite{cohen2020tensorlog} are two other neuro-symbolic and logic-oriented programming languages that aim to integrate neuronal and symbolic reasoning. Both rely on probabilistic logic and thus try to integrate neural primitives as predicates with probabilities that can be parameterized by neural networks. Similarly, Scallop \cite{huang2021scallop} takes a weighted graph approach for knowledge graphs, where the weights between entities are learned. Finally, SQLFlow \cite{wang2020sqlflow} focuses not on execution but on orchestration (using Kubernetes) for ML workloads using an SQL-like language. In our case the integration between neuronal and symbolic reasoning targets execution and occurs at a higher semantic level than probabilistic approaches. In NeuroQL reasoning tasks such as neuronal extraction and matching are expressed as filtering and binding operations for unification \cite{robinson1965machine,robinson2001handbook}. This allows us to create a concise neuro-symbolic query language that can act as a target for automatic translation of natural language questions.

A number of DSLs for neuro-symbolic programming \cite{chaudhuri2021neurosymbolic} target specific domains such as symbolic regression \cite{cranmer2020discovering}, behavior analysis \cite{sun2021task} and program synthesis \cite{ellis2020dreamcoder}. Ours is the first to address the problem of inter-subjective reasoning using neural translation, extending prior work on metadata \cite{ni2019justifying} and subjectivity \cite{bjerva2020subjqa}. 

Finally, our work is related to the wider domain of retrieval augmented  \cite{lewis2020retrieval,singh2021end,hstack2023} and tool augmented \cite{schick2023toolformer,lchain2023,levine2022standing} extraction and generation of answers. In our case though we proposed a rich intermediate representation for our task that takes the form of neuro-symbolic code. This allowed us to integrate neuronal and symbolic reasoning in an interpretable and explicit way.   

\section{Discussion} \label{sec:discussion}

We believe that the translation of natural language into neuro-symbolic instead of simply symbolic code is a step forward in the direction of closing the loop between neuronal and symbolic reasoning. The tasks, dataset and DSL for inter-subjective reasoning that we have introduced in this work provide a case-study close to real-world needs with clear neuro-symbolic characteristics. As such we believe it can be used as a baseline to evaluate future work.

Our results regarding hypothesis (\textbf{\textit{H1}}) for the \textit{NeuroQL Translation Task}, were surprising. For this task the network needs to learn not only to distinguish between objective and subjective components, but also between different kinds of sub-query categories (\ie \textit{opinion, title, price, reviews, manufacturing} \etc) as shown in Figure \ref{fig:translation}. Given the near perfect sacreBleu score for this task, we believe that the close-world assumption (\ie the fact that all queries concerned products and their objective or subjective properties) needs to be relaxed in future iterations of the dataset. For example we can consider extending the dataset with sufficiently different domains of inquiry, such as for \eg news articles and their commentary. Moreover, automatically paraphrasing \cite{fu2019paraphrase} our questions to create a larger more diverse dataset is a promising direction for future work. 

Our results regarding hypothesis (\textbf{\textit{H2}}) for the \textit{NeuroQL Query Task} were closer to our expectations given that there were comparable to a simpler extractive Q\&A task on user reviews reported by \cite{tunstall2022natural}. While these results can serve as a baseline, there is clearly more work needed to improve accuracy.

Regarding possible threats to validity, we note that the language used in parts of our questions comes directly from real-world usage and is unedited. This at first glance is a strength of the dataset (\ie models need to learn to adapt to real-world noise). Nevertheless, examples where there are syntactic or semantic mistakes (\eg from non-native speakers) can also affect the quality of labeling, making it harder to pinpoint valid answers. Here also, an automatic paraphrasing \cite{fu2019paraphrase} and auto-correction approach might help us control for and measure the impact of linguistic coherence in our dataset.

\section{Methods} \label{sec:methods}

\paragraph{\textit{Domain-Specific Language:}} We implemented NeuroQL as a DSL \cite{fowler2010domain,papoulias2019lands} using reflection and meta-programming \cite{maes1987concepts,papoulias2017end}, allowing us to change the normal semantics of Python only for NeuroQL expressions (leaving the rest of the runtime unchanged). These changes include \textit{(i)} the redefinition of variable declaration (when using the \lstinline[style=mystyle2]{vars()} and \lstinline[style=mystyle2]{ids()} primitives) \textit{(ii)} the redefinition of the dot, assignment and equality operators for NeuroQL variables and NeuroQL ids. These changes make NeuroQL an \textit{embedded pidgin} language according to the categorization of DSLs provided in \cite{renggli2010dynamic}.

\paragraph{\textit{Unification:}} Our base unification algorithm \cite{robinson1965machine,robinson2001handbook} is implemented as a generalization of pattern matching \cite{abelson1996structure,hanson2021software}, where both the pattern and the target may contain variables. The generalization receives a set of bindings as input (usually referred to as \textit{frames}) and returns possible augmentations that contain matched values of -- as yet -- unresolved variables. By allowing for sub-queries to be matched either jointly or in parallel we can implement logical operations such as conjunction and disjunction. Unification with inference rules can then be achieved through nested pattern matching that tries to satisfy a rule's body given the current bindings. 

\paragraph{\textit{BM25:}} We use a slight variation of the BM25 (Best Match 25) algorithm in our work to compensate for the most common cases of the ``exact overlapp'' problem of the initial algorithm \cite{jurafsky2023speech}. The BM25 is itself an improvement over TF-IDF (Term Frequency-Inverse Document Frequency) \cite{jurafsky2023speech}, with both algorithms being used as ranking methods for sparse retrievers. In our case we pre-process both the documents and the query to be matched by tokenizing and stemming the inputs. We don't provide a further solution for synonyms using BM25, since in this case the more accurate DPR primitive can be used (see below). The two methods (BM25 and DPR) provide a trade-off for our users between speed and accuracy, with BM25 being the fastest and DPR being the most accurate.

\paragraph{\textit{Dense Passage Retrieval:}} DPR (Dense Passage Retrieval) is a method that uses dense text embeddings for both query and documents that need to be matched. It is based on a BERT bi-encoder architecture that computes a dot product similarity between a document and a query. Our DPR is based on \cite{karpukhin2020dense}, using question \& context encoders trained with the Natural Questions dataset \cite{lee-etal-2019-latent,kwiatkowski-etal-2019-natural}. This DPR is then used as a backend for of our \textit{\lstinline[style=mystyle2]{neural_match}} primitive. The \textit{\lstinline[style=mystyle2]{neural_match}} primitive receives a sub-query whose results are used to create a set of \textit{(id, document)} pairs. It then tries to match the documents against a target query text, using our DPR to return the top\_k results that matched (see full example of \textit{\lstinline[style=mystyle2]{neural_match}} in Section \ref{sec:neuro-symbolic-composition}).

\paragraph{\textit{Reader Model:}} %
A Reader or reading comprehension model is a neural network that can perform extractive Q\&A by returning relevant text intervals of documents. In our work we use the MiniLM \cite{wang2020minilm} model initially trained on the SQuAD 2.0 dataset \cite{rajpurkar2018know} and further fine-tuned on the NeuroQL training set to improve its performance (as we describe in Section \ref{sec:experiments_and_validation}). This Reader is then used as a backend for of our \textit{\lstinline[style=mystyle2]{neural_extract}} primitive. The \textit{\lstinline[style=mystyle2]{neural_extract}} primitive receives the name of a new variable to bind for extracted answers, a query to create \textit{(id, document)} pairs and finally a target query text. It then tries to extract relevant text intervals from our documents using the Reader to return the top\_k results found. We fine-tuned our reader model for \textit{3 epochs}, using a \textit{384}-token sequence length with a \textit{128}-token document stride. We used a batch size of \textit{16}, learning rate of \textit{1e-5} with a \textit{0.2} warmup and included the ability to return no answers when predicting results.

\paragraph{\textit{Translation Model:}} A translation model is a sequence-to-sequence neural network trained over pairs of input and target sequences. In our case we fine-tuned a CodeT5 model \cite{wang2021codet5}) using the NeuroQL question/query dataset in order to translate inter-subjective questions posed in natural language into the NeuroQL query that can answer them (as detailed in Sections \ref{sec:neuroql_dataset} and \ref{sec:experiments_and_validation}). This translation model is then used as a backend for our \lstinline[style=mystyle2]{NeuroQL.translate} and \lstinline[style=mystyle2]{NeuroQL.answer} primitives, which as their name suggests can translate questions into NeuroQL code and attempt to answer them.
Our translation model was fine-tuned for \textit{20 epochs} with a maximum input/output length of \textit{512} tokens, a batch size of \textit{64} and an \textit{1e-4} learning rate using a \textit{linear} scheduler and an \textit{AdamW} optimizer \cite{loshchilov2017decoupled}.
 
\paragraph{\textit{SacreBleu, Recall, EM \& F1 Scores:}} In machine translation tasks the \textbf{\textit{sacreBleu}} \cite{post2018call} score is used for the evaluation of generated translations focusing on reproducibility and comparability of reported results. It is itself a standardization of the Bleu \cite{papineni2002bleu} score that compares the n-grams of the generated sequences to those of the reference translations. In our work we use \textit{sacreBleu} to evaluate the  quality of our translation model (as detailed in Section \ref{sec:experiments_and_validation}). \textbf{\textit{Recall:}} is a metric used to evaluate retrieval methods such as our \textit{\lstinline[style=mystyle2]{neural_match}} primitive that is using a DPR model. Recall represents the percentage of relevant documents retrieved among the top\_k results returned by a retriever (as we report in Section \ref{sec:experiments_and_validation}). \textbf{\textit{EM \& F1 Scores:}} are used to evaluate Q\&A extraction methods such as our \textit{\lstinline[style=mystyle2]{neural_extract}} primitive that is using a Reader model. EM represents the percentage of exact extracted matches, while F1 measures the harmonic mean of precision and recall (reported in Section \ref{sec:experiments_and_validation}).

\section*{Code \& Data Availability} \label{sec:code_and_data_availability}

All data and examples from this paper are available at: \url{https://orgdlabs.com/neuroQL}

\bibliography{neuroQL}

\end{document}